\documentclass{article}
\begin{document}
\title{Chern--Simons States in\highlighting{SO(1,n)}
  Yang--Mills Gauge Theory of Quantum Gravity}
\title{Chern--Simons States in SO(1,n)  Yang--Mills Gauge Theory of
Quantum Gravity}

\newcommand{\orcidauthorA}{0000-0000-0000-000X} 

\author{Zbigniew Haba\\ Institute of Theoretical Physics,
University of Wroclaw,\\email:zbigniew.haba@uwr.edu.pl}
\maketitle









\begin{abstract}We discuss a quantization of the Yang--Mills theory
with an internal symmetry group $SO(1,n)$ treated as a unified
theory of all interactions. In one-loop calculations, we show that
Einstein gravity can be considered as an approximation to gauge
theory. We discuss the role of the Chern-Simons wave functions in
the quantization.
\end{abstract}




\section{Introduction}
We discuss a quantization of the Yang--Mills theory with   an
internal symmetry $SO(1,n)$. When some matter fields are added,
 then the model can be considered as a candidate for a unified theory of
all interactions. The conceptual
origin of such an approach to unification came from the paper of
Yang \cite{yang}. Soon it was followed by Cho \cite{cho1},
Freund~\cite{cho2}, MacDowell, and Mansouri~\cite{mansouri}. The non-compact internal gauge group also appears in models of supergravity \cite{nieuv,au}.
The $SO(1,n)$ model with $n\geq 13$ ($n=13$ in \cite{cham}, $n=17$ \mbox{in \cite{zoup}} for string-inspired higher-dimensional theories ) could encompass gravity within the Georgi--Glashow
\cite{georgi} or Pati--Salam \cite{pati}   extensions of the Standard Model (SM) based on
$SO(10)$ internal symmetry group. It is shown in \cite{cham} that the Brout--Englert--Higgs symmetry-breaking mechanism will break $SO(1,13)$ symmetry to $SO(1,3)\times SU(3)\times SU(2)\times U(1)$
in agreement with the SM phenomenology. $SO(1,17)$  is needed (according to~\cite{zoup})
in order to obtain a correct content of Weyl--Majorana spinors in models resulting from higher-dimensional superstring theories.
In such a model, the $so(1,n)$ connection is decomposed into the $so(1,3)$ Lorentz connection,
the vierbein and the compact gauge fields, which are supposed (after  symmetry breaking)
to describe weak and strong interactions of SM.
The vierbein is used to describe the metric. Then, both the vierbein and the $so(1,3)$ connection are dynamical
variables. The decomposition seems ad hoc. There is a scheme where the metric appears as a result of  $SO(1,4)$ or  $SO(2,3)$ symmetry breaking by  scalar fields \cite{west,wilczek2} (see also \cite{guendel}).
 With the vierbein-connection decomposition,
it has been noticed early in the study of the  $SO(1,n)$ model \cite{townsend}
that the Yang--Mills Lagrangian can be decomposed into the (non-compact) $SO(1,3)$ Yang--Mills part, the
Einstein--Hilbert Lagrangian (in the Palatini form), the cosmological term, and the $SO(n-4)$ (compact) Yang--Mills Lagrangian. Unfortunately, there is an arbitrariness in the coupling constants in this decomposition. Without a coupling to matter fields, the values of these constants (crucial for the phenomenology) cannot be fixed. In addition, the bare constants undergo
renormalization (depending on the matter field content) in quantum theory.
The project encounters many other problems in the composition of matter fields in the
realization of SO(n-4) symmetry breaking for an
agreement with the phenomenology of particle physics (for recent reviews, see \cite{zoup,procc}).
Concerning the weak  interactions, the crucial CP violation is described by a set of left and
right Weyl fermions (mirror fermion problem \cite{mirror}). The problem of the composition of fermions in  SM may be related to gravity, as discussed \mbox{in \cite{nesti,alex1,alex2}}. The formulation of classical and quantum gravity in Ashtekar variables \cite{ashtekar,smolin,samuel,nieto,chopin}  uses (anti)self-dual connections which are not invariant under spatial reflections. Their coupling to fermions violates parity
 invariance (on the other hand, determinants of Weyl fermions give parity non-invariant Chern--Simons terms
 \cite{wittenrev,redlich}).
A calculation of the effective action from a coupling of the   $SO(1,3)$ connection  to fermions
can lead to new terms in the model of weak interactions.

   The
Yang--Mills theory with a non-compact symmetry group
\cite{townsend,alex2,don1,don2,don3,don4} is
renormalizable, asymptotically free, and infrared-unstable. The
infrared instability can be interpreted as a confinement of the
$SO(1,4)$ connection \cite{don2} , which could justify the classical
limit of the Yang--Mills theory to its Einstein  form. There are
some difficult problems in the Yang--Mills interpretation of
gravity. The non-compactness of the internal symmetry group has as
a consequence that the energy is not bounded from below. This does
not lead to difficulties in the Feynman integral approach to
quantum Yang--Mills theory. However, the unitarity of the model
is questionable (see a discussion of a possible solution of the problem in~\cite{alex2}). An approach based on the  Euclidean
 formulation does not answer this question.
The analytic continuation of $SO(1,n)$ to $SO(n+1)$ allows us to
quantize the model on the lattice in the Euclidean framework
\cite{seiler}, but the continuation to Minkowski space-time will
lead to difficulties. In the canonical quantization, the  negative
energy could be interpreted as the gravitational energy, but this
interpretation needs further investigation. There is still the
question of the general coordinate invariance and the equivalence
principle in gauge theories of gravity, which require an
explanation. In three-dimensional gravity, the gauge invariance and
the general coordinate invariance can be unified \cite{witten0}.

 In this paper, we concentrate on the gauge field self-interaction.
 In sec.2
we show that if the Lagrangian is represented as a
 quadratic form plus the Chern--Simons divergence term; then,
 a Chern--Simons wave function $\psi_{CS}$ does not change in time; i.e., it is an
 eigenstate. As a consequence, the Schr\"odinger equation of the wave function
 $\psi=\psi_{CS}\chi$ can be expressed as a diffusion equation for $\chi$ with an (anti)self-dual drift.
 Then, in the leading order of the $\hbar$ expansion, the solution of the Schr\"odinger equation is expressed by the solution of the self-duality equation. The higher orders can be obtained from a solution
 of a stochastic self-duality equation. In Section~\ref{s3}, we explain the method in the Abelian model.
 The Chern--Simons (CS) state is not invariant under the spatial reflection
 ${\bf x}\rightarrow -{\bf x}$.
~Such states may appear from an interaction with Weyl fermions in physical models of weak interactions, but we do not discuss such models
 (see \cite{nesti,alex2}). We calculate correlation functions in CS states.
The CS wave function is not normalizable \cite{witten}. We show in Abelian gauge theories that a rigorous calculation of correlation functions with a regularized CS wave function has a limit that can be considered as an analytic continuation of formal Gaussian calculations~\cite{witten1,witten2,witten3}.
Such states are of interest in condensed matter physics in 2 + 1 dimensions (anyons~\cite{wilczek},
charged vortices with fractional statistics \cite{frohlich}). We expect that they may be relevant in
models with gravitational and weak interactions \cite{alex2}. The Chern--Simons Lagrangian
defines a topological field theory related to quantum gravity \cite{rovelli}.

  Our main concern in this paper is the problem of the undue number of
 variables in the $SO(1,n)$ Yang--Mills model (Section~\ref{s4}) as compared with the Palatini form of the
 Einstein--Hilbert action. The $SO(1,n)$ Yang--Mills Lagrangian contains both the vierbeins and
 the $SO(1,3)$ connection as dynamical variables. It is suggested in \cite{don1,don2,don3}
 that, owing to the infrared confinement, the connection being confined does not play any dynamical role.
 In the standard Palatini approach, the functional integration over the Lorentz connection
 leads to the Einstein--Hilbert action. The procedure is equivalent to the expression of the
 connection in terms of vierbeins on the basis of classical equations of motion. In Section~\ref{s5}, we perform the functional integral in the $\hbar$ (loop) expansion  over the connection in the $SO(1,4)$
 Yang--Mills model.
 As a result, we obtain the effective action, which consists of the classical action and the quantum
one-loop  correction. We show that if the square of the curvature
of the Lorentz $SO(1,3)$ connection is neglected, then until the
first order in the expansion in $\hbar$, we obtain the classical
Einstein action (in Palatini form) plus the one-loop quantum
correction of order $\hbar$. The result shows that  in the loop expansion, we can consistently control
the Palatini (infrared) approximation and the ultraviolet behavior of the model.

In Section~\ref{s6} we discuss an expansion in $\hbar$
in  the $SO(1,n)$ Yang--Mills model by a stochastic perturbation of the self-duality
equation. Then, the  correlation functions of the Wilson loops can
be calculated in regularized CS states in a perturbation
expansion.

\section{A Total Divergence Term in the
Path Integral }\label{s2}
 The time evolution of the state
$\psi(A)$ dependent on the variable $A$ describing the system is
given by the path integral formula (integration is over paths
$A_{s}(A)$ starting from $A$, and $\psi$ is the initial condition for
the Schr\"odinger equation)
\begin{equation}
\psi_{t}(A)=\int {\cal
D}A\exp\Big(\frac{i}{\hbar}\int_{0}^{t}d{\bf x}ds{\cal
L}(A_{s})\Big)\psi(A_{t}(A)).
\end{equation}
We assume 
 that the Lagrangian  can be expressed in the form
\begin{equation}
L=\int d{\bf x}{\cal L}=\int d{\bf x}(\frac{1}{2}Q(A){\cal
M}Q(A)+\partial_{\mu}J^{\mu}),
\end{equation}where ${\cal M}$ does not depend on $A$.
Then, the path integral (1) reads (we assume that  $\int d{\bf
x}\nabla{\bf J}=0$)
\begin{displaymath}
\begin{array}{l}
\psi_{t}(A)=\exp\Big(-\frac{i}{\hbar}\int d{\bf
x}J^{0}(A)\Big)\int {\cal
D}A\exp\Big(\frac{i}{\hbar}\int_{0}^{t}d{\bf
x}ds\frac{1}{2}Q(A){\cal
M}Q(A)\Big)\cr\times\exp\Big(\frac{i}{\hbar}\int d{\bf
x}J^{0}(A_{t}(A))\Big)\psi(A_{t}(A)).\end{array}
\end{displaymath}
Let \begin{displaymath} W=i\int d{\bf x}J^{0}(A).
\end{displaymath}
We express the initial state $\psi$ in the form
\begin{equation}
\psi=\exp(-\frac{1}{\hbar}W)\chi\equiv\psi_{CS}\chi.
\end{equation}  Then\begin{equation}\begin{array}{l}
\psi_{t}(A)=\exp(-\frac{W}{\hbar})\int {\cal
D}A\exp\Big(\frac{i}{\hbar}\int_{0}^{t}d{\bf
x}ds\frac{1}{2}Q(A_{s}){\cal
M}Q(A_{s})\Big)\chi(A_{t}(A))\cr\equiv
\exp(-\frac{W}{\hbar})\chi_{t}(A).\end{array}
\end{equation}
If the Jacobian $Z^{-1}=\frac{\partial Q}{\partial A}\simeq
const$, then we can change variables $ A \rightarrow Q(A)$ so that
\begin{displaymath}
\chi_{t}(A)=\Big< \chi (A_{t}(Q))\Big>,
\end{displaymath}
where $Q$ is a Gaussian variable. We can express this change of
variables as
\begin{equation}
Q(A_{s})=\sqrt{i\hbar}\partial_{s}B_{s},
\end{equation}
where $B_{s}$ (the Brownian motion) has Gaussian distribution with
the functional measure
\begin{displaymath}
{\cal D}B\exp\Big(-\frac{1}{2}\int ds(\partial_{s}B)^{2}\Big).
\end{displaymath}

If we choose $\chi=1$ then $\psi=\exp(-\frac{1}{\hbar}W)$ and
Equation~(4) reads
\begin{equation}
\begin{array}{l}
\psi_{t}(A)=\psi(A)\int {\cal
D}A\exp\Big(\frac{i}{\hbar}\int_{0}^{t}d{\bf
x}ds\frac{1}{2}Q(A){\cal M}Q(A)\Big).\end{array}
\end{equation}
Hence, $\psi_{t}(A)=Z\psi(A)$ for $\psi=\exp(-\frac{W}{\hbar})$,
where the factor $Z$ does not depend on $A$. If the Jacobian
$\frac{\partial Q}{\partial A}\simeq const$ (as will be the case
in our models), then the integral on the RHS of Equation~(6) is a time-independent constant $\det{\cal M}^{-\frac{1}{2}}$.
  It
follows from Equation~(6) that $\psi=\exp(-\frac{1}{\hbar}W)$ does not
change over time.

The formula (3) can  be considered either as a similarity
transformation of states in quantum mechanics or as a result of a
change of the canonical formalism resulting from the presence of an
extra time derivative in the Lagrangian
\cite{habaepjplus,habarxiv}. In both interpretations, the
modified Schr\"odinger equation for the Lagrangian quadratic in
velocity reads
\begin{equation}
i\hbar\partial_{t}\chi=\frac{1}{2}\Pi\Pi\chi+((\psi_{CS})^{-1}\Pi\psi_{CS})\Pi\chi,
\end{equation}where $\Pi$ is the quantum realization of the
canonical momentum.
\section{Quantization of an Electromagnetic Field on a Lorentzian Manifold}\label{s3}
We consider first the Lagrangian of the electromagnetic field
\begin{equation}\begin{array}{l}
{\cal
L}=-\frac{1}{4}\sqrt{-g}g^{\mu\nu}g^{\alpha\beta}F_{\mu\alpha}F_{\nu\beta}
=-\frac{1}{8}\sqrt{-g}g^{\mu\nu}g^{\alpha\beta}(F_{\mu\alpha}\pm
\frac{i}{2}\tilde{\epsilon}_{\mu\alpha\sigma\rho}F^{\sigma\rho})(F_{\nu\beta}\pm
\frac{i}{2}\tilde{\epsilon}_{\nu\beta\sigma\rho}F^{\sigma\rho})\cr\mp
\frac{i}{4}\epsilon_{\mu\alpha\sigma\rho}F^{\mu\alpha}F^{\sigma\rho}
=-\frac{1}{8}\sqrt{-g}F^{\pm}_{\mu\alpha}F^{\pm\mu\alpha}\mp\frac{i}{4}F^{*}_{\mu\alpha}F^{\mu\alpha}
\end{array}
\end{equation}
defined on a Lorentzian manifold with the metric $g_{\mu\nu}$ ($g$
denotes the determinant of the metric). In Equation~(8),
\begin{equation}
\tilde{\epsilon}_{\mu\alpha\sigma\rho}=\sqrt{-g}\epsilon_{\mu\alpha\sigma\rho},\end{equation}
where on the LHS we have the tensor density $\tilde{\epsilon}$ and
on the RHS we have the Levi-Civita antisymmetric symbol on the Minkowski
space-time defined by $\epsilon^{0123}=1$,
$F^{*}_{\mu\nu}=\epsilon_{\mu\nu\alpha\beta}F^{\alpha\beta}$.

The last term in the Lagrangian (8) is of the form (2) as
\begin{equation}
\frac{i}{2}\epsilon_{\mu\alpha\sigma\rho}F^{\mu\alpha}F^{\sigma\rho}=i\partial_{\mu}
(A^{\alpha}\epsilon_{\mu\alpha\sigma\rho}F^{\sigma\rho})\equiv
i\partial_{\mu}J^{\mu}.
\end{equation}The self-dual variable $F^{\pm\mu\alpha}$ can be chosen as
the Gaussian variable $Q$. As a consequence of Equation~(6), the
Chern--Simons state
\begin{equation}
\psi_{CS}=\exp(\pm\frac{1}{2\hbar}\int d{\bf
x}A_{j}\epsilon^{jkl}\partial_{k}A_{l})
\end{equation}
is an eigenstate. It does not change in time because the Jacobian
$\frac{\partial Q}{\partial A}\simeq \exp\int \frac{\partial
Q}{\partial A\partial A}=const$. The term $\partial_{\mu}J^{\mu}$
does not depend on the metric (it is a topological invariant, the
\mbox{Chern--Simons term}).

We prove $H\psi_{CS}=0$, independently constructing in the
canonical formalism the Hamiltonian $H$ for the quantum
electromagnetic field on a globally hyperbolic manifold. Then,
 we can choose coordinates \cite{global} such that the metric takes
the form
\begin{equation}
ds^{2}=g_{00}dtdt-g_{jk}dx^{j}dx^{k} \end{equation} (we use Greek
letters for space-time indices and Latin letters for spatial
indices). Then, the canonical momentum in the $A^{0}=0$ gauge is

\begin{equation}
\Pi^{j}=g^{00}\sqrt{-g}g^{jk}\partial_{0}A_{k}.
\end{equation}
The Hamiltonian is
\begin{equation}
H=\frac{1}{2}\int\frac{g_{00}}{\sqrt{-g}}g_{jk}\Pi^{j}\Pi^{k}+\frac{1}{4}\int\sqrt{-g}g^{jk}g^{ln}F_{jl}F_{kn}
\end{equation}
with the canonical momentum
\begin{equation}
\Pi^{j}=-i\hbar\frac{\delta}{\delta A_{j}}.
\end{equation}
We prove that on a globally hyperbolic manifold in the metric (12),
$H\psi_{CS}=0$. This \mbox{condition requires}
\begin{equation}\begin{array}{l}
\frac{1}{2}\int
g_{00}(-g)^{-\frac{1}{2}}g_{jk}\Pi^{j}\Pi^{k}\psi_{CS}=- \int
g_{00}(-g)^{-\frac{1}{2}}g_{jk}\epsilon^{jrl}\epsilon^{kmn}F_{rl}F_{mn}\psi_{CS}\cr
=-\frac{1}{4}\int
\sqrt{-g}g^{jl}g^{rm}F_{jr}F_{lm}\psi_{CS}\end{array}
\end{equation}

The general proof  needs some strenuous calculations
\cite{habarxiv}. It is simple in the case of the orthogonal
isotropic metric
\begin{equation}
ds^{2}=a_{0}^{2}dt^{2}-a^{2}d{\bf x}^{2}.
\end{equation}
 In this metric on the LHS of Equation~(16) \begin{equation}\begin{array}{l} \int
g_{00}(-g)^{-\frac{1}{2}}g_{jk}\Pi^{j}\Pi^{k}\psi_{CS}=-\int
a_{0}a^{-1}\epsilon^{jrl} F_{rl}\epsilon^{jmn} F_{mn}\psi_{CS}
\end{array}
\end{equation}
whereas on the RHSk we have
 \begin{equation}\begin{array}{l}\frac{1}{2}\int
\sqrt{-g}g^{jl}g^{rm}F_{jr}F_{lm}\psi_{CS} =\int
a_{0}a^{-1}F_{kl}F_{kl}.\end{array}
\end{equation}
Hence, (18) is equal to (19).

Using the transformation (4), we obtain the Schr\"odinger equation
for $\chi$ (a version of Equation~(7)):
\begin{equation}
\partial_{t}\chi=\int d{\bf x} \Big(\frac{i\hbar}{2}a_{0}a^{-1}\frac{\delta^{2}}{\delta
{\bf A}({\bf x})^{2}}\pm
ia_{0}a^{-1}\epsilon_{jkl}\partial_{k}A_{l}\frac{\delta}{\delta
A_{j}({\bf x})}\Big)\chi.
\end{equation}If the term $\hbar\frac{\delta^{2}}{\delta
{\bf A}({\bf x})^{2}}$ in the limit $\hbar\rightarrow 0$ is
neglected, then the solution of Equation~(20) is expressed by the
solution ${\bf A}_{t}({\bf A})$ (with the initial condition ${\bf
A}$) of the self-duality equation
\begin{equation}\begin{array}{l}
F^{(\pm)}_{\mu\nu}=\pm\frac{i}{2}\tilde{\epsilon}_{\mu\nu\alpha\beta}F^{\pm\alpha\beta}=
\pm\frac{i}{2}\epsilon_{\mu\nu\alpha\beta}\sqrt{-g}F^{\pm\alpha\beta}.
\end{array}\end{equation}
Then, \begin{equation} \chi_{t}({\bf A})= \chi({\bf A}_{t}({\bf
A}))
\end{equation}
(written for a time-independent metric).

The transformation to Gaussian variables $Q$ discussed in Equation~(5)
can be expressed as a stochastic equation corresponding to the
$\psi_{CS}$ state. In the time-independent isotropic orthogonal
metric, it reads (see
refs.~\cite{habaepj,habaentropy,hababook,simon}
for a general framework)
\begin{equation}\begin{array}{l}
dA_{j}(s)=\pm ia_{0}a^{-1}\epsilon_{jkl}\partial_{k}A_{l}ds+
\sqrt{\frac{i\hbar a_{0}}{a}}dB_{j}(s),\end{array}\end{equation}
where the covariance of the Brownian motion is
\begin{equation}
E[B_{j}(t,{\bf x})B_{k}(s,{\bf
y})]=min(t,s)(\delta_{jk}-\partial_{j}\partial_{k}\triangle^{-1})
\delta({\bf x}-{\bf y}).\end{equation}
  Then, Equation~(22) (after  averaging over
the Brownian motion) determines an exact solution of Equation~(20). The
solution of the self-duality Equation (21) gives the leading order
behavior of $\chi$. The corrections can be calculated by means of
an averaging over the Brownian motion.

The Chern--Simons wave function $\psi_{CS}$ is not square-integrable. We need $\chi$ in \mbox{Equation~(4)} to decay sufficiently fast in
order to make $\psi$ square-integrable. It turns out that  $\psi $
may preserve some crucial properties of $\psi_{CS}$ after a
multiplication by $\chi$ (treated as a regularization). As an
example, we consider QED on the Minkowski space-time with
\begin{equation}
\chi=\exp(-\frac{\gamma}{2}({\bf A},\omega{\bf A})),
\end{equation}where (,) denotes the $L^{2}$ scalar product, $\omega=\sqrt{-\triangle}$ and
$\partial_{j}A_{j}=0$.
 We calculate the
correlation functions (the integral over ${\bf A}$ is well-defined
for $\gamma\hbar>1$) at $t=0$ in the state $\psi=\psi_{CS}\chi$.
We obtain
\begin{equation}
\begin{array}{l}(\psi,A_{j}({\bf p})A_{k}({\bf
p}^{\prime})\psi)=\hbar\delta({\bf p}+{\bf p}^{\prime})(
\frac{\hbar\gamma}{\hbar^{2}\gamma^{2}-1}(\delta_{jk}-p_{j}p_{k}p^{-2})
+\frac{i}{\hbar^{2}\gamma^{2}-1}\epsilon_{jrk}p^{r}p^{-2}).
\end{array}\end{equation} Equation (26) is singular at
$\gamma\hbar=1$. It makes sense at $\hbar\gamma=0$. We may treat
the RHS of Equation~(26) for $\gamma\hbar<1$ as an analytic continuation
in the complex $\gamma$ plane of the integral on the LHS of
Equation~(26). A definition of the Gaussian integral  of a function
non-integrable in the Riemann sense by means of an analytic
continuation appears in the ordinary Feynman oscillatory integrals
with the Chern--Simons Lagrangian
\cite{witten1,witten2,witten3}.

We calculate the loop expectation values in the regularized CS
states for the loops $C_{1}$ and $C_{2}$
\begin{equation}\begin{array}{l}
(\psi,\int_{C_{1}} {\bf A}d{\bf x}_{1} \int_{C_{2}} {\bf A}d{\bf
x}_{2}\psi)= -\hbar 4\pi link(C_{1},C_{2})+{\cal
O}(\hbar\sqrt{\hbar}),
\end{array}\end{equation} where $link(C_{1},C_{2})$ denotes the Gauss linking number.
It follows from Equation~(22) that after the time evolution, we
obtain the linking number up to the terms of order
$\hbar\sqrt{\hbar}$.

\section{\boldmath{$SO(1,n)$} Yang--Mills Gauge Theory on a Lorentzian Manifold }\label{s4}
We consider an $o(1,n) $-valued  connection 1-form
\begin{equation}
\Omega_{\mu}=h_{\mu}^{a\alpha}J_{a\alpha}+\omega_{\mu}^{ab}J_{ab}+\omega^{\alpha\beta}J_{\alpha\beta}\equiv
\Omega_{\mu}^{AB}J_{AB},
\end{equation}
where  $J_{AB}$ with $A,B=0,1,\dots,n$ are the generators of
rotations in $AB$ planes of the $O(1,n)$ group, $J_{AB}=-J_{BA}$ ,
$a,b=0,1,2,3$ , $\alpha=4,\dots, n$, and $J_{\alpha\beta}$ are the
generators of $SO(n)$ rotations. We have
\begin{displaymath}
[J_{AB},J_{CD}]=\eta_{AD}J_{BC}+\eta_{BC}J_{AD}-\eta_{AC}J_{BD}-\eta_{BD}J_{AC}
\end{displaymath}where $\eta_{AB}$ is the Minkowski metric $(-,+,\dots,+)$ in n + 1 dimensions.

 We may introduce a dimensional parameter $\sigma$ rescaling $h\rightarrow \sigma h$. This is equivalent to a
  rescaling of the $J_{a5}$ generators
 ($J_{a4}\rightarrow \sigma J_{a4}$) in the
commutators $J_{AB}$ so that
\begin{equation} [J_{a4},J_{b4}]= \sigma^{2}J_{ab},
\end{equation}
where $J_{ab}$ satisfy the standard commutation relations of the
Lorentz group for the rotations in the $ab$ plane. When
$\sigma\rightarrow 0$, then $J_{a4}$ will contract to the
generators of momenta $P_{a}$ of the Poincare group.
 The curvature of
the connection $\Omega$ can be defined by the commutator of
covariant derivatives $D_{\mu}=\partial_{\mu}+\Omega_{\mu}$
\begin{displaymath}
[D_{\mu},D_{\nu}]={\cal R}_{\mu\nu}={\cal R}_{\mu\nu}^{AB}J_{AB},
\end{displaymath}
\vspace{-12pt}
\begin{equation}
{\cal R}^{AB}_{\mu\nu}=\partial_{\mu}\Omega^{AB}_{\nu}
-\partial_{\nu}\Omega^{AB}_{\mu}+f^{AB}_{CD;MN}(\Omega_{\mu}^{CD}
\Omega_{\nu}^{MN}-\Omega_{\nu}^{CD} \Omega_{\mu}^{MN}),
\end{equation}where $f^{AB}_{CD;MN}$ are structural constants of
the group $O(1,n)$ (after the rescaling (29)).

For notational simplicity, we restrict our discussion to $n=4$.
$n>4$ adds the compact group $SO(n-4)$ , which is supposed to
describe internal symmetries. We decompose \mbox{${\cal
R}^{AB}_{\mu\nu}$ as}\vspace{-6pt}
\begin{equation}\begin{array}{l} {\cal
R}^{ab}_{\mu\nu}=\partial_{\mu}\omega_{\nu}^{ab}-\partial_{\nu}\omega_{\mu}^{ab}
+\omega^{ac}_{\mu}\omega^{db}_{\nu}\eta_{cd}-\omega^{ac}_{\nu}\omega^{db}_{\mu}\eta_{cd}
+\sigma^{2}(h^{a}_{\mu}h^{b}_{\nu}-h^{b}_{\mu}h^{a}_{\nu}).
\end{array}\end{equation} For the term ${\cal
R}_{\mu\nu}^{a4}J_{a4}\equiv R^{a}_{\mu\nu}J_{a4}$, we have
\begin{equation}\begin{array}{l}
R^{a}_{\mu\nu}=\partial_{\mu}h^{a}_{\nu}-\partial_{\nu}h^{a}_{\mu}+\omega_{\mu}^{am}\eta_{mr}h^{r}_{\nu}-
\omega_{\nu}^{am}\eta_{mr}h^{r}_{\mu}=(D_{\mu}h)_{\nu}^{a}-(D_{\nu}h)_{\mu}^{a}\equiv
{\cal T}^{a}_{\mu\nu},\end{array}
\end{equation}where $(D_{\mu}h)_{\nu}^{a}=\partial_{\mu}h^{a}_{\nu}
+\omega_{\mu}^{mr}\eta_{mr}h^{r}_{\nu}$, $\eta_{ab}=(-1,+1,+1,+1)
$ is the Minkowski metric and ${\cal T}$ is the torsion .

We consider as the non-Abelian generalization of Equation~(8) the
Lagrangian of the $SO(1,4)$ Yang--Mills field
\begin{equation}\begin{array}{l}
{\cal
L}=-\frac{1}{4\lambda^{2}}\sqrt{-g}g^{\mu\nu}g^{\alpha\beta}{\cal
R}^{AB}_{\mu\alpha}{\cal R}^{CD}_{\nu\beta}g_{AB;CD}
\cr=-\frac{1}{8\lambda^{2}}\sqrt{-g}g^{\mu\nu}g^{\alpha\beta}({\cal
R}^{AB}_{\mu\alpha}\pm
\frac{i}{2}\tilde{\epsilon}_{\mu\alpha\sigma\rho}{\cal
R}^{\sigma\rho;AB})({\cal R}^{CD}_{\nu\beta}\pm
\frac{i}{2}\tilde{\epsilon}_{\nu\beta\sigma\rho}{\cal
R}^{\sigma\rho;CD})g_{AB;CD}\cr\mp
\frac{i}{4\lambda^{2}}\epsilon_{\mu\alpha\sigma\rho}{\cal
R}^{\mu\alpha:AB}{\cal R}^{\sigma\rho;CD}g_{AB;CD}
\equiv-\frac{1}{8\lambda^{2}}\sqrt{-g}{\cal
R}^{\pm}_{\mu\alpha;AB}{\cal
R}^{\pm\mu\alpha;AB}\mp\frac{i}{4\lambda^{2}}{\cal R}^{*
AB}_{\mu\alpha}{\cal R}^{\mu\alpha}_{AB}
\end{array}
\end{equation}
defined on a Lorentzian manifold with the metric $g_{\mu\nu}$. In
Equation~(33), $
\tilde{\epsilon}_{\mu\alpha\sigma\rho}=\sqrt{-g}\epsilon_{\mu\alpha\sigma\rho}$
as in Equation~(9), where on the LHS, we have the tensor density
$\tilde{\epsilon}$ , ${\cal
R}^{*}_{\mu\nu}=\epsilon_{\mu\nu\alpha\beta}{\cal
R}^{\alpha\beta}$. $\lambda^{2}$ is the dimensionless coupling
constant for the Yang--Mills theory. $g_{AB;CD}=Tr(J_{AB}J_{CD})$ is
the Killing form on the algebra $o(1,n)$.

The representation of the Lagrangian (33) leads to the
Chern--Simons term
\begin{equation}
\frac{i}{2\lambda^{2}}\epsilon_{\mu\alpha\sigma\rho}{\cal
R}^{\mu\alpha}{\cal R}^{\sigma\rho}= i\partial^{\mu}J_{\mu}.
\end{equation}
 Let us note that
$\partial_{\mu}J^{\mu}$ does not depend on the metric (it is a
topological invarian, the Chern--Simons term). It has the form
\begin{displaymath}
J_{0}=\frac{1}{\lambda^{2}}Y_{CS},
\end{displaymath}
where
\begin{displaymath}
Y_{CS}=-\int d{\bf x}Tr(\frac{1}{2}\Omega
d\Omega+\frac{1}{3}\Omega\wedge\Omega\wedge\Omega)
\end{displaymath}Specifically, in components in $SO(1,4)$,
\begin{equation}\begin{array}{l}
Y_{CS}=\epsilon_{jkl}\Big(\frac{1}{2}h_{j}^{a}\partial_{k}h_{l}^{b}\eta_{ab}+\frac{1}{2}\omega_{j}^{ab}h_{k}^{c}h_{l}^{d}\eta_{ac}\eta_{bd}
\cr+\frac{1}{2}\omega_{j}^{ab}\partial_{k}\omega_{l}^{cd}\eta_{ac}\eta_{bd}
+\frac{1}{3}\eta_{ra}\eta_{pb}f^{rp}_{cd;mn}\omega_{j}^{ab}\omega_{k}^{cd}\omega_{l}^{mn}\Big),
\end{array}\end{equation}
where $f^{ab}_{cd;mn}$ are the structure constants of $o(1,3)$.
 If $h_{\mu}^{a}$ is interpreted as the tetrad (vierbein), then we can see  that the Lagrangian (33) contains
  the Einstein--Hilbert Lagrangian (in the Palatini formalism) as well as the  Yang--Mills part
(the square of the $SO(1,3)$ field strength), the cosmological
term, and  the square of the torsion. In fact, expanding the
Lagrangian (33), we get
\begin{equation}\begin{array}{l}
{\cal L}=-\frac{1}{4\lambda^{2}}\sqrt{-g}g^{\mu\nu}g^{\alpha\beta}
R^{ab}_{\mu\alpha}
R^{cd}_{\nu\beta}\eta_{ac}\eta_{bd}-\frac{\sigma^{2}}{2\lambda^{2}}\sqrt{-g}g^{\mu\nu}g^{\alpha\beta}
R^{ab}_{\mu\alpha}(h_{\nu}^{c}h^{d}_{\beta}-h_{\nu}^{d}h^{c}_{\beta})\eta_{ac}\eta_{bd}
\cr-\frac{\sigma^{4}}{4\lambda^{2}}\sqrt{-g}g^{\mu\nu}g^{\alpha\beta}(h_{\nu}^{c}h^{d}_{\beta}-h_{\nu}^{d}h^{c}_{\beta})
(h_{\mu}^{a}h^{b}_{\alpha}-h_{\mu}^{b}h^{a}_{\alpha})\eta_{ac}\eta_{bd}
\cr-\frac{\sigma^{2}}{4\lambda^{2}}\sqrt{-g}g^{kj}g^{lm}((D_{k}h)_{l}^{a}-(D_{l}h)_{k}^{a})
((D_{m}h)_{j}^{b}-(D_{j}h)_{m}^{b}),\end{array}
\end{equation}
where
$(D_{k}h)^{a}_{l}=\partial_{k}h^{a}_{l}+\omega_{k}^{ab}h_{l}^{b}$.
Till now, in Equation~(33), we did not specify the manifold and the metric $g_{\mu\nu}$ on this manifold. The unification of gravity and gauge theory is based on a choice of the metric,
identifying the $J_{a4}$ part of the $so(1,n)$ connection (28) with the vierbein
\cite{cham,zoup,townsend}. It can be seen from Equation~(36) that  the choice
\begin{equation}
g^{\mu\nu}=h^{\mu}_{a}\eta^{ab}h_{b}^{\nu},
\end{equation}
where $h^{\mu}_{a}$  satisfy
\begin{equation}
\eta^{ab}h^{\mu}_{b}h_{\mu}^{c}=\eta^{ac}
\end{equation}
leads in Equation~(36) to the appearance (as the second term) of the
Einstein Lagrangian in the Palatini formalism, if we choose
$\sigma^{2}=\frac{\lambda^{2}}{8\pi G}=\lambda^{2}m_{PL}^{2}$, where $m_{PL}$ is the Planck mass.
 The term
four-linear in $h$ in Equation~(36) is the cosmological term. The last
term is a square of the torsion (32). As a consequence of the identification (37),
the $SO(1,4)$ symmetry is broken. The Hamiltonian depends on the Ansatz (37).
The Chern--Simons wave function does not depend on the metric. As follows from Secs.2-3, the insertion of the metric (37) does not change the conclusion that
$\psi_{CS}$ does not evolve in time (it is an eigenfunction of the Hamiltonian).
The identification of the connection $h_{\mu}^{a}$  with the vierbein in the metric (37)
has already been discussed  by Townsend  \cite{townsend} in  classical theory.
In  quantum theory (as discussed in Section~\ref{s5}), all four terms are separately renormalized.
 The bare Yang--Mills coupling  $\lambda$ tends to zero at small distances
 (asymptotic freedom), but it can become large at large distances
 (confinement of the connection; see
\cite{don1,don2,don3,don4}). When discussing physical consequences of the choice of bare couplings and renormalization,
we should consider the $SO(1,n)$  gauge field model as immersed in a unified GUT
model. In such a case, the renormalization is changed by the couplings to other fields (with a preservation of the asymptotic freedom).
The cosmological constant $\frac{\sigma^{4}}{\lambda^{2}}$  acquires a large (positive) contribution depending on the matter field content.

\section{Functional Integration Over the Lorentz Connection}\label{s5}
We perform the functional integral (1) over $\omega$ (assuming that
$\psi$ does not depend on $\omega$ ) in an expansion in $\hbar$.
We obtain as a result the factor $\exp\Big(\frac{i}{\hbar}{\cal
L}_{eff}(h)\Big)(1+O(\sqrt{\hbar}))$ defining the effective
action. We treat $h_{\mu}^{a}$ as an external field and decompose
$\omega=A+\sqrt{\hbar}\tilde{\omega}$ into the classical part $A$
and quantum fluctuations $\tilde{\omega}$ as in the background
field method. We write
\begin{equation}
\Omega_{\mu}^{CD}J_{CD}=(A_{\mu}^{CD}+h_{\mu}^{CD})J_{CD}+\sqrt{\hbar}\hat{\omega}_{\mu}^{ab}J_{ab}
\equiv \tilde{A}_{\mu}+\sqrt{\hbar}\hat{\omega},
\end{equation}where $A^{AB}_{\mu}$ has only the $ab$ components
and $h_{\mu}^{AB}$ has only $a4$ components. We expand the
Lagrangian (36) in $\sqrt{\hbar}\hat{\omega}$. In order to cancel
(in the expansion) the terms proportional to $\sqrt{\hbar}$, we
choose $\tilde{A}$ in such a way that
\begin{equation}\begin{array}{l}
(\tilde{\nabla}^{\mu}(\tilde{A}){\cal R}_{\mu\nu}(\tilde{A}))^{ab}
= \nabla^{\mu}\delta_{ab;cd}{\cal R}^{cd}_{\mu\nu}+
f^{ab}_{CD;EF}\tilde{A}^{\mu CD}{\cal R}^{EF}_{\mu\nu}= 0
\end{array}\end{equation} This is a non-linear equation for
$A^{ab}_{\mu}$. We can solve it by iteration. In the lowest order,
(linear in $A_{\mu}^{ab}$), only $f^{ab}_{4c;4e}$ terms contribute
to the last term in Equation~(40).  Then, Equation~(40) reads (with the term
quadratic in $A$ neglected)\begin{equation}\begin{array}{l}
\frac{1}{\sqrt{-g}}\partial^{\mu}\sqrt{-g}\Big(\partial_{\mu}A_{\nu}^{ab}
-\partial_{\nu}A_{\mu}^{ab}+\sigma^{2}(h_{\mu}^{a}h^{b}_{\nu}
-h_{\nu}^{b}h^{a}_{\mu})\Big) \cr +h^{\mu
a}(\partial_{\mu}h^{b}_{\nu}-
\partial_{\nu}h^{b}_{\mu}+A_{\mu}^{bc}h_{\nu}^{c}-A_{\nu}^{bc}h_{\mu}^{c})
+h^{\mu a}\Gamma^{\rho }_{\nu\mu}h^{b}_{\rho}\eta_{ab}=0.
\end{array}\end{equation}
When the first term (coming from $\partial^{\mu}R^{ab}_{\mu\nu}$)
is neglected, then we obtain the equation for \mbox{zero torsion}
\begin{equation}\begin{array}{l}
h^{\mu a}\Big(\partial_{\mu}h^{b}_{\nu}-
\partial_{\nu}h^{b}_{\mu})+A_{\mu}^{bc}h_{\nu}^{c}-A_{\nu}^{bc}h_{\mu}^{c}.
+\Gamma^{\rho }_{\nu\mu}h^{b}_{\rho}\eta_{ab}\Big)=0
\end{array}\end{equation}
This equation has the solution \cite{schwinger}
\begin{equation}
A_{abc}=\frac{1}{2}(G_{bca}+G_{cab}-G_{abc}),
\end{equation}
where $A_{abc}=h_{a}^{\mu}A_{\mu bc}$, $G_{cab}=h_{\mu
c}G^{\mu}_{ab}$ and
\begin{equation}
G^{\mu}_{ab}=h^{\nu}_{a}\partial_{\nu}h_{b}^{\mu}-h^{\nu}_{b}\partial_{\nu}h_{a}^{\mu}.
\end{equation}
It follows that by neglecting the derivative term (the first term in
(41)) in the functional integration over $\omega$, we obtain in the
leading order of the $\hbar$ expansion the Einstein gravity. This
could also be seen as a consequence of neglecting the first
(Yang--Mills) term in the Yang--Mills action (36) and taking into
account only the second (Palatini) term. In an exact $SO(1,4)$
quantum Yang--Mills theory at the one-loop level, we obtain in the
functional integral the classical term
$\exp(\frac{i}{\hbar}L(\tilde{A}))$ (the tree approximation) and
subsequently the one-loop contributions $\det(
\tilde{\nabla}^{2}g_{\mu\nu}+2i{\cal R}_{\mu\nu})^{-\frac{1}{2}}$
and the Fadeev--Popov determinant, whose form depends on the choice
of gauge (see the calculations in \cite{habaprd}). The
determinants must be renormalized.  It can be seen that at
one loop, the $O(1,4)$ gauge theory is asymptotically free
\cite{don1,don2,don3,don4}.

Specifically, we have
\begin{equation}\begin{array}{l}
{\cal R}^{AB}_{\mu\nu}(\tilde{A}+\sqrt{\hbar}\hat{\omega})={\cal
R}^{AB}_{\mu\nu}(\tilde{A})+
\sqrt{\hbar}(\nabla_{\mu}\hat{\omega}^{AB}_{\nu}-\nabla_{\nu}\hat{\omega}^{AB}_{\mu}
\cr+f^{AB}_{CD;MN}(\hat{\omega}_{\mu}^{CD}\tilde{A}_{\nu}^{MN}-\hat{\omega}_{\nu}^{CD}\tilde{A}_{\mu}^{MN}))
+O(\sqrt{\hbar}).\end{array}\end{equation} It follows that until
the terms of order $\hbar$,
\begin{equation}\begin{array}{l}
\int {\cal L}(\tilde{A}+\sqrt{\hbar}\hat{\omega})=\int {\cal
L}(\tilde{A})+ \sqrt{\hbar}\int\sqrt{-g}(\tilde{\nabla}^{\mu}{\cal
R}_{\mu\nu})^{AB}\hat{\omega}^{CD}g_{AB;CD}\cr+\frac{\hbar}{2}\int
\sqrt{-g}\hat{\omega}^{\mu
AB}\Big(-\tilde{\nabla}_{\alpha}\tilde{\nabla}^{\alpha}g_{\mu\nu}g_{AB;CD}+R_{\mu\nu}(h)g_{AB;CD}+
\tilde{\nabla}_{\mu}^{AB}\tilde{\nabla}^{CD}_{\nu}\cr+2f^{EF}_{AB;CD}{\cal
R}_{\mu\nu EF }(\tilde{A})\Big)\hat{\omega}^{\nu
CD},\end{array}\end{equation} where $R_{\mu\nu}(h)$ is the Ricci
tensor of the metric (37) defined by the tetrad $h$ and
\begin{displaymath}
(\tilde{\nabla}_{\mu}\hat{\omega}_{\nu})^{AB}=\partial_{\mu}\hat{\omega}^{AB}_{\nu}
+f^{AB}_{CD;MN}\tilde{A}_{\mu}^{CD}\hat{\omega}_{\nu}^{MN}+\Gamma_{\nu\mu}^{\beta}\hat{\omega}^{AB}_{\beta}
\end{displaymath}
The requirement of a vanishing of the $\sqrt{\hbar}$ term leads to Equation~(40). There
remains
\begin{displaymath}
\int {\cal L}=\int {\cal
L}(\tilde{A})+\frac{\hbar}{2}\int\hat{\omega}{\cal M}\hat{\omega}.
\end{displaymath}

 The
integration over $\hat{\omega}$ gives the determinant $\det{\cal
M}^{-\frac{1}{2}}$. It is convenient to choose the $\xi$ gauge
leading to the factor \begin{equation}
\exp\Big(-\frac{1}{2\xi}Tr\int
\sqrt{-g}(\tilde{\nabla}_{\mu}\hat{\omega}^{\mu})^{2}\Big).
\end{equation}
In particular, if $\xi=1$,  then the
$\tilde{\nabla}_{\mu}\tilde{\nabla}_{\nu}$ term in Equation~(46) is
vanishing. The determinant of ${\cal M}$ can be represented by
means of a Feynman--Kac formula as in \cite{habaprd}. In the gauge
(47), the Fadeev--Popov determinant is equal to the determinant of
the  Laplacian for vector fields
$-\tilde{\nabla}_{\alpha}\tilde{\nabla}^{\alpha}g_{\mu\nu}g_{AB;CD}+R_{\mu\nu}(h)g_{AB;CD}$.
We can conclude that for the renormalization of the determinant
${\cal M}$, we need the terms
\begin{equation}\begin{array}{l}
\int\Big(c_{1}\sqrt{-g}g^{\mu\nu}g^{\alpha\beta}{\cal
R}_{\mu\alpha}^{AB}(\tilde{A}){\cal
R}_{\nu\beta}^{CD}(\tilde{A})g_{AB;CD} +c_{2}\sqrt{-g}R^{2}(h)
\cr+c_{3}\sqrt{-g}R_{\mu\nu}(h)R^{\mu\nu}(h)+c_{4}\sqrt{-g}\Big),
\end{array}\end{equation} where $R^{\mu\nu}(h)$ is the Ricci tensor for the metric (37) and $R(h)$ is the scalar curvature.
 In the effective action, there still will  be the logarithm of the
Faddev--Popov determinant  whose renormalization requires the same
counterterms as in Equation~(48) (with different constants). In Equation~(48),
besides the  square of the $SO(1,4)$ curvature of the connection
$\Omega$ appearing in the renormalization of the Yang--Mills theory
on a flat manifold (the first term), we have the counterterms required for the
renormalization of the determinant of the Laplacian
$g_{\mu\nu}\nabla^{2}+R_{\mu\nu}$ for the vector fields on a
curved background. Then, the $SO(1,4)$ gauge invariance of the
model is broken as a consequence of the identification (37).

 The one-loop expansion is easily generalized to $SO(1,n)$.
We shall have the coupling of the $S0(1,4)$ gauge field to the
$SO(n-4)$ Yang--Mills as in an $SO(1,4)\times O(n-4)$ Yang--Mills
theory. If the $o(n-4)$ algebra is denoted as $L_{\alpha\beta}$
Then we just make a change in Equation~(28) for the background field
expansion $\tilde{A}+\sqrt{\hbar}\hat{\omega}\rightarrow
\tilde{A}+Y^{\alpha\beta}L_{\alpha\beta}+\sqrt{\hbar}\hat{\omega}$
Where $Y_{\mu}$ is the $O(n-4)$ gauge field. It is also possible
to achieve a non-trivial extension of internal and space-time
symmetries by constructing the Yang--Mills theory with the group
$O(p,q)$ .
\section{Chern--Simons States}\label{s6}
The evolution Equation (7) for $SO(1,4)$ theory is a
generalization of the one for the Hamiltonian (14). In the gauge
$\Omega_{0}=0$, Equation~(7) reads
\begin{equation}\begin{array}{l}
i\hbar\partial_{t}\chi=\int
\frac{1}{2}g^{00}\frac{1}{\sqrt{-g}}g_{jk}(\Pi^{jAB}
\Pi^{kCD}g_{AB;CD} +2g_{AB;CD}(\Pi^{jAB}Y_{CS})\Pi^{kCD})\chi.
\end{array}\end{equation}
 In the quantized model,\begin{equation}
\Pi^{jCD}=-i\hbar\frac{\delta}{\delta \Omega_{j}^{CD}}
\end{equation}or in components,
\begin{equation}
\Pi^{ja}=-i\hbar\frac{\delta}{\delta h_{j}^{a}}
\end{equation}and
\begin{equation}
\Pi^{jab}=-i\hbar\frac{\delta}{\delta \omega_{j}^{ab}}.
\end{equation}
In the Hamiltonian evolution Equation (49), we need
\begin{equation}\begin{array}{l}
\Pi^{jCD}Y_{CS}=i\hbar\epsilon^{jkl}{\cal
R}_{kl}^{CD}=i\hbar\epsilon^{jkl}\Big(\partial_{k}\Omega_{l}^{CD}-\partial_{l}\Omega_{k}^{CD}
+f^{CD}_{GH;EF}\Omega_{k}^{GH}\Omega_{l}^{EF}\Big)
\end{array}\end{equation}In components
\begin{equation}
\Pi^{ja}Y_{CS}=i\hbar\epsilon^{jkl}R_{kl}^{a}=i\hbar\epsilon^{jkl}\Big(\partial_{k}h_{l}^{a}-\partial_{l}h_{k}^{a}
+\omega^{ab}_{k}h^{b}_{l}-\omega^{ab}_{l}h^{b}_{k}\Big)
\end{equation}
and
\begin{equation}\begin{array}{l}
\Pi^{jab}Y_{CS}=i\hbar\epsilon^{jkl}{\cal
R}_{kl}^{ab}=i\hbar\epsilon^{jkl}\Big(\partial_{k}\omega_{l}^{ab}-\partial_{l}\omega_{k}^{ab}
\cr+\omega^{ac}_{k}\omega_{l}^{db}\eta_{cd}-\omega^{ac}_{l}\omega_{k}^{db}\eta_{cd}
+\sigma^{2}(h^{a}_{k} h^{b}_{l}-h^{b}_{k} h^{a}_{l})\Big)
\end{array}\end{equation}
It follows already from the change of variables (5) that in the
leading order of $\hbar$ expansion of the solution of the
Schr\"odinger equation with the initial condition $\psi_{CS}\chi$
is
\begin{equation}
\chi_{t}(\Omega)=\chi(\Omega_{t}(\Omega))+{\cal O}(\sqrt{\hbar}),
\end{equation}
where $\Omega_{t}(\Omega)$ is the solution of the self-duality
equation (with the initial condition $\Omega$)
\begin{equation}
{\cal R}_{\mu\nu}^{AB}=i\epsilon_{\mu\nu\alpha\beta}{\cal
R}^{\alpha\beta AB}\sqrt{-g},
\end{equation}
i.e.,
\begin{equation}
{\cal R}_{0j}^{AB}=i\epsilon_{jkl}g^{km}g^{ln}{\cal
R}_{mn}^{AB}\sqrt{-g}.
\end{equation}
The self-duality equation
 in the gauge $\Omega_{0}=0$ can be
expressed as \begin{equation}\begin{array}{l}
\partial_{t}\Omega^{AB}_{j}=i\epsilon_{jkl}g^{km}g^{ln}\sqrt{-g}(\partial_{m}\Omega^{AB}_{n}-
\partial_{n}\Omega^{AB}_{m}+f^{AB}_{CD;EF}\Omega_{m}^{CD}\Omega_{n}^{EF}).
\end{array}\end{equation}
As the $\Pi\Pi$ part in the Schr\"odinger Equation (49) is of
a higher order in $\hbar$, we confirm on the basis of the
calculations (53)--(55) that the time evolution (56) generated by
the \mbox{classical flow}
\begin{equation}
g^{00}\frac{1}{\sqrt{-g}}g_{jk}
g_{AB;CD}(\Pi^{kAB}Y_{CS})\frac{\delta}{\delta \Omega_{j}^{CD}}
\end{equation}
is the solution of the Schr\"odinger Equation (49) in the leading
order in $\hbar$. In order to derive an exact solution or
calculate  higher orders in $\hbar$, we need a stochastic equation
(a generalization of Equation~(23)).

As mentioned in the introduction, the Chern--Simons states arise
from the total derivative terms in the Lagrangian, which appear in
Yang--Mills as well as in other gauge theories  of gravity (see
\cite{nieto,chopin}). The wave functions  $\psi_{CS}$ are
not normalizable. One can regularize these states in order to make
them normalizable while still preserving their exceptional
properties~\cite{habaepjplus,habarxiv}. After the removal of
the regularization, the result is the same as a formal (Gaussian)
functional integral. In this way, we can perform a perturbative
calculation of the Wilson loop variables
\begin{equation}
U_{C}=TrP(\exp i\int_{C} \Omega_{\mu}dx^{\mu}).
\end{equation}
We could consider the correlation functions in the $\psi_{CS}$
``ground state''
\begin{equation}\begin{array}{l}
<U_{C_{1}}\dots.U_{C_{n}}>=\int \psi_{CS}^{2}U_{C_{1}}\dots.U_{C_{n}}.
\end{array}\end{equation}
In the lowest order of calculations, we obtain
\begin{equation}\begin{array}{l}
<U_{C_{1}}\dots.U_{C_{n}}>=1+4\pi
\hbar\sum_{kl}link(C_{k},C_{l})+\dots
\end{array}\end{equation}
where $link(C_{k},C_{l})$ is the linking number of the curves
$C_{k}$ and $C_{l}$ and the higher order terms involve expectation
values of powers of $\Omega$  and higher orders of $\hbar$. The
result (63) is a consequence of the formula
\begin{equation}\begin{array}{l}
\int {\cal D}\omega\exp\Big( \int
\omega^{j}\epsilon_{jkl}\partial^{k}\omega^{l}
\Big)\omega_{m}({\bf x})\omega_{n}({\bf
y})\cr=\epsilon_{rmn}(x_{r}-y_{r})\vert {\bf x}-{\bf y}\vert^{-3}
\end{array}\end{equation} and a similar formula for the tetrads
$h_{j}^{a}$.

$Y_{CS}$ (35) is trilinear in $\Omega$ with the propagator
$\partial^{-1}$. By elementary power counting, a perturbation
expansion for a theory with the Lagrangian $Y_{CS}$ is
renormalizable in \mbox{three dimensions}. This follows from the formula for
the divergence index $\nu$ \cite{hababook,ryder} for
theories with a propagator $p^{-1}$ as
$\nu=d-E\frac{d-1}{2}+V(n\frac{d-1}{2}-d)$ where $d$ is the
dimension, $n$ is the degree of the interaction polynomial, $E$ is
the number of external lines, and $V$ is the number of vertices. In
our Chern--Simons  model, $\nu=d-E$ . It follows that we can
calculate the correlation functions in a perturbation theory in
the Chern--Simons states.

\section{Summary and Outlook}\label{s7}
We have discussed the special role of Chern--Simons wave functions in
Yang--Mills theories. By means of a similarity transformation, they
transform the Schr\"odinger equation into a diffusion-type
equation with the (imaginary) diffusion constant proportional to
$\hbar$ and an $\hbar$-independent drift, which is a functional
derivative of the Chern--Simons term. As a consequence, a solution
of the Schr\"odinger equation in the leading order of $\hbar$ is
expressed by the solution of the self-duality equation. The exact
solution of the Schr\"odinger equation can be expressed by the
solution of a stochastic equation, which is a perturbation by noise
of the self-duality equation. We have applied this framework to
the Yang--Mills theory based on an
internal $SO(1,n)$ symmetry group proposed by  some authors. We
discussed a quantum version of these models in an $\hbar$
expansion. We have derived in  detail the one-loop formula for the
effective action in $SO(1,n)$ gauge theories, showing that at some
limit, the quantum version contains the Einstein gravity
interacting with the Yang--Mills connection. We discussed the
Chern--Simons wave function and its role in the calculation of
correlation functions of Wilson loop variables. The considered wave functions (which
are a perturbation of the Chern--Simons wave functions) are of a special form. We believe that
states of this form can arise in models describing an interaction of gravity with
gauge theories and matter fields. In particular, an effective action arising from weak interactions can lead to Chern--Simons terms.
  In a prospective
Study there remain interesting questions about the consequences of
the asymptotic freedom of quantum gauge theories at small
distances and the infrared instability (confinement) at large
distances for the gravitational part of the $SO(1,n)$ Yang--Mills Lagrangian.
The content of matter fields of GUT models could determine the couplings for $SO(n-4)$
symmetry breaking and possibly the CS-type states describing the CP violation in weak interactions.
Then, the time evolution of such states (including their scattering amplitudes) could be determined
by a solution of the self-duality equations. The gravity itself in the unified model would be described by states
invariant under diffeomorphisms (e.g., certain diffeomorphism-invariant square integrable regularizations of the CS wave functions).
\vspace{6pt}

\end{document}